\documentclass[preprint,12pt]{elsarticle}

\usepackage{amssymb}

\usepackage{colortbl} 
\usepackage{xcolor}
\usepackage{soul}
\usepackage{amsmath}
\usepackage[utf8]{inputenc}

\usepackage{fontenc, inputenc, calc, indentfirst, fancyhdr, graphicx, lastpage, ifthen, lineno, float, amsmath, setspace, enumitem, mathpazo, booktabs, titlesec, etoolbox, tabto, xcolor, soul, multirow, totcount,changepage, attrib, upgreek, hyphenat, natbib, hyperref, url, geometry, newfloat, ulem, colortbl}

\usepackage{algorithm,algpseudocode}
 
\journal{arXiv}

\begin{document}

\begin{frontmatter}



\title{ZTCloudGuard: Zero Trust Context-Aware Access Management Framework to Avoid Misuse Cases in the Era of Generative AI and Cloud-based Health Information  Ecosystem}






\author[first]{Khalid Al-hammuri}
\author[first]{Fayez Gebali}
\author[second]{Awos Kanan}
\affiliation[first]{organization={University of Victoria},Electrical and Computer Engineering
            addressline={Finnerty Rd.}, 
            city={Victoria},
            postcode={V8W 2Y2}, 
            state={BC},
            country={Canada}}

            \affiliation[second]{organization={Princess Sumaya University for Technology}, Computer Engineering
            addressline={Khalil Saket Street Al-Jubaiha}, 
            city={Amman},
            postcode={11941}, 
            state={Amman},
            country={Jordan}}
\begin{abstract}
 Managing access between large numbers of distributed medical devices has become a crucial aspect of modern healthcare systems, enabling the establishment of smart hospitals and telehealth infrastructure. However, as telehealth technology continues to evolve and Internet of Things (IoT) devices become more widely used, they are also becoming increasingly exposed to various types of vulnerabilities and medical errors. In healthcare information systems, about 90\% of vulnerabilities emerged from misuse cases and human errors. As a result, there is a need for additional research and development of security tools to prevent such attacks. This article proposes a zero-trust-based context-aware framework for managing access to the main components of the cloud ecosystem, including users, devices and output data. The main goal and benefit of the proposed framework is to build a scoring system to prevent or alleviate misuse cases while using distributed medical devices in cloud-based healthcare information systems.
The framework has two main scoring schemas to maintain the chain of trust. First, it proposes a critical trust score based on cloud-native micro-services of authentication, encryption, logging, and authorizations. Second, creating a bond trust scoring to assess the real-time semantic and syntactic analysis of attributes stored in a healthcare information system. The analysis is based on a pre-trained machine learning model to generate the semantic and syntactic scores. The framework also takes into account regulatory compliance and user consent to create a scoring system. The advantage of this method is that it is applicable to any language and adapts to all attributes as it relies on a language model, not just a set of predefined and limited attributes. The results show a high $F1$ score of 93.5\%, which proves that it is valid for detecting misuse cases.
\end{abstract}




\end{frontmatter}


\section{Introduction}\label{sec:introduction}

\textbf{Problem statement:}
The article proposes a zero-trust access management framework for the healthcare information system. The article also conducted a case study on misuse cases to verify the framework's viability. \\
Misuse cases in health care are defined as circumstances that lead to a wrong medical decision. The decision may lead to prescribing the wrong drug \cite{cousins2023prescription,islam2023artificial}, issuing the wrong report, or a false diagnosis.  The misuse is very critical as it may be caused by normal users, not just by fraud. In healthcare  90\% of vulnerabilities are from misuse cases or human errors. This is difficult to detect as it is caused by authorized and authenticated users. In the era of generative AI assistance, which could generate biased, discriminated, or even wrong medical reports \cite{volovici2022steps}.  The misuse cases are caused either by human or AI-assistant systems. The proposed framework is designed to make a scoring system for any type of user, data or device.

Controlling access to devices and their users (either human or AI) and associated data in healthcare systems is a major challenge for any service provider.  While promoting the idea of smart hospitals and telehealth, it is required to look deeply into the existing regulation and access control systems to be more valid in the context of using technologies like internet of things (IoT) devices, cloud \cite{cloud_nancy2022iot}, AI \cite{AI_valizadeh2022ai,AI_Loh_2022}, blockchain \cite{blockchain_Chauhan_2022, iomt_blockchain_lakhan2022federated}, quantum computing \cite{quantum_rasool2022,quantum2_kumar2022} and 5G networks \cite{5gchen2020security}.
There are different types of medical devices that are used within the distributed or cloud-based healthcare information environment. Examples of these devices are patient monitoring devices, handheld and portable devices, telehealth consulting, medical imaging systems, robotics, and virtual reality.\\

The main challenges facing the cloud-based healthcare infrastructure are managing access to these devices and guaranteeing that the received data is secure, clean, and clinically valid.  The stressful environment and complex technology require serious attention and advanced skills to operate such systems. At the same time, the devices and the data should be monitored in real-time to intercept any abnormalities in data or wrong reports sent by healthcare practitioners. Such a system should control users, data and output.

\textbf{Constraints:}
The healthcare industry is subject to strict regulations regarding the use of patient information. In the United States, the Health Insurance Portability and Accountability Act (HIPAA) governs patient information compliance. Similarly, in Canada, Bill (C-27) regulates the healthcare information system, and Health Canada also plays a role. Bill (C-27) is a new law that replaces The Personal Information Protection and Electronic Document Act (PIPEDA) and has enacted three regulations: the Consumer Privacy Protection Act (CPPA), the Personal Information and Data Protection Tribunal Act (PIDPTA), and the Artificial Intelligence and Data Act (AIDA). In Europe, compliance with the General Data Protection Regulation (GDPR) regulates the sharing of information in healthcare.

\textbf{Proposed solution:}
To adapt to the new advancement in technology and strict compliance, the paper proposes a zero-trust context-aware access control framework for medical IoT devices that can manage the patient information system within the complex structure of healthcare systems. The framework proposes a set of practices, policies and attributes to enhance and manage the security of access control systems for any healthcare infrastructure. The framework mainly focuses on preventing misuse cases in the healthcare industry. These cases are complex and difficult to detect as they are sent by authorized and authenticated users. The proposed context-aware system can alleviate the user's errors by analyzing the complex metadata of the user, device and output. The proposed context-aware framework ensures that the data is relevant,  consistent, authenticated and only sent by authorized users to the designated destination at the end point device or user.\\

The contributions of this article are as follows:
\begin{itemize}
    \item Propose zero trust context-aware access management framework that relies on the user ($x$), hardware ($y$) and output data ($z$), Sec. \ref{sec.proposedframework}.
    \item Design critical trust assessments based on cloud-native micro-services of authentication, authorization, encryption and logging, Sec. \ref{sec.trustsore}.
    \item  Implement bond trust scoring assessment based on the analysis of syntactic and semantic relationship between $x$, $y$ and $z$. Bond trust is used to build a chain of trust between different entities in healthcare information systems. It is derived using a pre-trained word2vec machine learning model, Sec. \ref{sec.trustsore}.
    \item Construct hierarchical encoding for decision-making of the framework to grant a complex final decision that considers regulatory compliance, access constraints, access level and access operations, Sec. \ref{sec.accesshierarchy}.
    \item Conduct a study to evalaute the framework to assess the validity of this framework on preventing misuse cases in healthcare industry. Sec. \ref{sec.result}.
    
\end{itemize}

The rest of the article is organized as follows. Sec. \ref{sec.background} provides a background of the current access control work. Sec. \ref{sec.proposedframework} explain the proposed framework in detail. Sec. \ref{sec.result} report and discuss the experiment results. Sec. \ref{sec.conclusion} concludes the paper.

\section{Background and current access control system for IoT devices in healthcare } \label{sec.background}

While the traditional access control systems are sufficient within the local hospital environment, they have limited security or functionality outside the perimeter of the local healthcare infrastructure. The emerging need for distributed IoT devices that require internet connections and advanced server or cloud infrastructure to store, compute and manage data interoperability makes it challenging, especially in the era of AI, which requires a new access control policy. The existing healthcare information systems rely on the HL7FHIR standards \cite{securityfhir} to define the communication and security access control system in healthcare. There are three main subsystems within the HL7FHIR:

\begin{enumerate}
    \item \textbf{Authentication:} To verify the user.
    \item \textbf{Access Control Engine:} To decide which FHIR controls are allowed for the user using the \textbf{CRUD} method (Create, Read, Update, Delete).
    \item \textbf{Audit log:} to record the actions and any suspicious system intrusion.
\end{enumerate}


At the organizational level, the access control system has three main common types within the health information system:
\begin{itemize}
 
    \item \textbf{RBAC:} Role-based access control \cite{role_zhang2022fine}, \cite{rashid2020securing}.
    \item \textbf{ABAC:}  Attribute-based access control \cite{attribute_khan2022efficient,attribute2_Sanders_2019}.
    \item \textbf{CML:} Modern cloud-based machine learning access control \cite{ml_nobi2022administration, ml_Nobi_2022, cloudml_Jin_2022}.
\end{itemize} 

The \textbf{RBAC} and \textbf{ABAC} are the standard access control systems in healthcare. They are used widely in the traditional healthcare infrastructure setup for managing access control within the hospital perimeter. The \textbf{RBAC}  manages the access based on the user role and grants permission based on the \textbf{CRUD} or \textbf{HTTP} method.  \textbf{RBAC} is complex and considers different factors like users (operator, patient), roles, permissions, resources objects and context of the data access \cite{outchakoucht2017dynamic, securityfhir}. Please refer to  Table \ref{tabaroleinfo} in Appendix \ref{aceessinfo} for more information on the Role-based access system factors. The limitation of a rule-based access management system is that it is time-consuming and requires a lot of manual work to adjust rules and policies, which makes it less effective in real-time access management that has too many factors in a complex cloud-based environment.

The \textbf{ABAC} \cite{chiquito2023attribute, hu2015attribute} grants the user permission to access the system resources or objects based on predefined policies and conditions that are characterized by specific data attributes. For example, to adhere to compliance regulations, the patient identification information can not be accessed without having the patient consent attribute to process the data.  The ABAC has a few challenges; it is difficult to scale, which limits its use in the cloud ecosystem. In addition, it needs a considerable amount of time and resources, which makes it inflexible and inefficient in a dynamic and modern globally distributed healthcare system. 

To access health information in a traditional access engine, the user typically sends a request to the server through the REST API gateway. The server then sends a request for the REST API to verify the user information to grant the required \textbf{CRUD} operations based on the predefined rules and policies.  Authenticating IoT devices also is challenging and requires special techniques. For example, the Physical Unclonable  Function (PUF) is used to authenticate the hardware for the telehealth distributed devices \cite{gebali2022sram}.

The modern Cloud-based ML access control systems rely mainly on the zero trust principle that analyzes everything in the network and does not grant trust to any entity, either a user or device, for data access without passing a set of conditions that are defined by the organization policy. However, the main challenge is to identify what attributes or data context that will be considered in the policy without compromising the quality of the provided service. Mitigating the risk within the network is also essential to evaluate the access decision of the users \cite{ghorbani2020systems}.

\begin{figure}[http]
\begin{center} 
\includegraphics[width=14 cm]{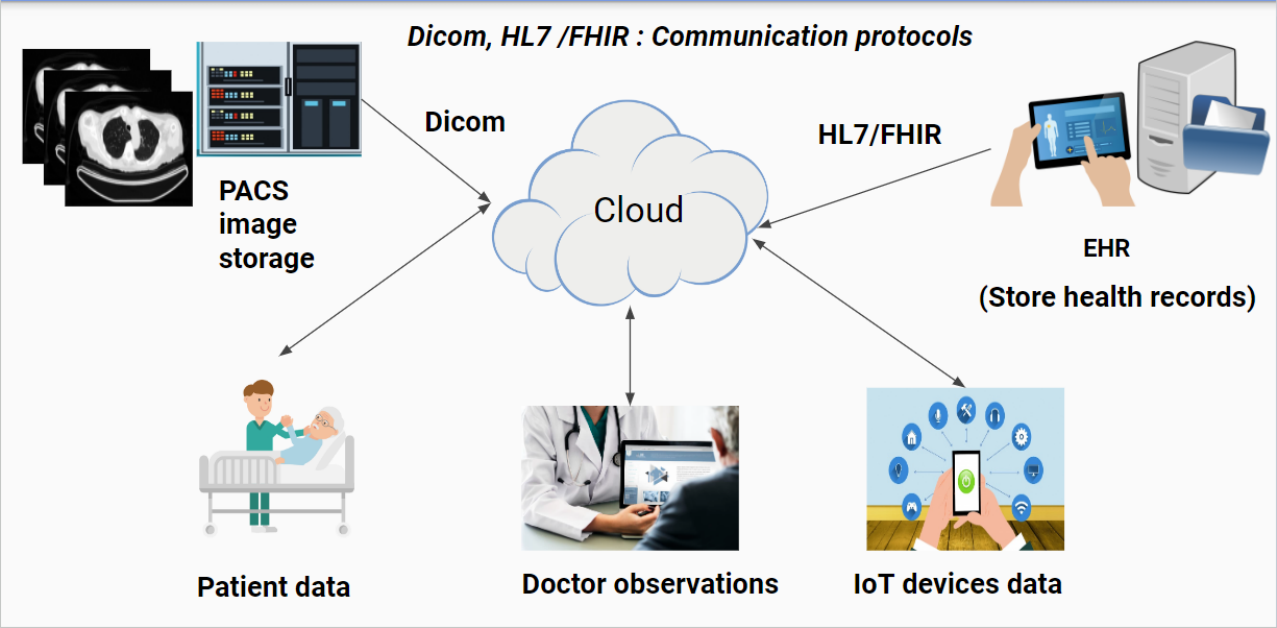}
\caption{Illustration of the main data sources of IoT devices in Cloud-based telehealth ecosystem. } \label{fig.datasource}
\end{center}
\end{figure}

Fig. \ref{fig.datasource} depicts the main sources of data for IoT devices in the cloud-based healthcare environment. Picture Archiving Communication System (PACS) stores images in Digital Imaging and Communications in Medicine (DICOM) format. Electronic Health Records (EHR) keep the patient records using High-Level Seven (HL7) and  Fast Healthcare Interoperability Resources (FHIR) standards. Patient monitoring data is also sent in real-time for data analysis, doctor observations, and data from different distributed IoT devices. These data sources are the essential backbone for the health information system data sources.  Data ingestion and processing in Digital health applications can be handled using AI tools like transformer \cite{Al_hammuri_2023_vit}. The transformer is a state-of-the-art AI algorithm that can extract the relationship between different healthcare components using an attention mechanism.

\section{Method}\label{sec.proposedframework}

\subsection{Overview of the proposed  zero trust framework for access management}
\begin{figure}[H]
\begin{center} 
\includegraphics[width=14 cm]{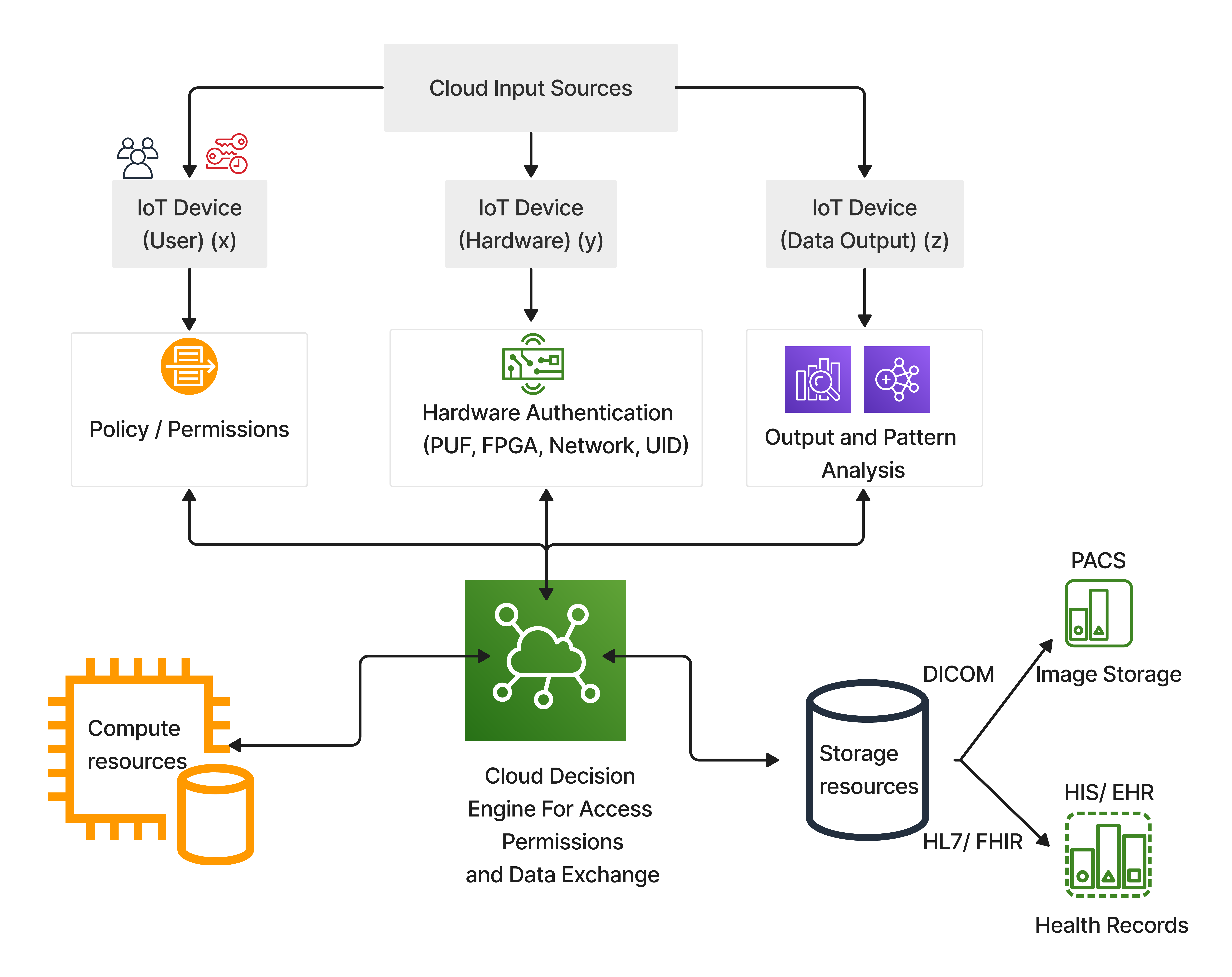}
\caption{Representative of the proposed access control functional diagram within the healthcare cloud-AI ecosystem.  } \label{fig.accessdiagram}
\end{center}
\end{figure}

The architecture design for the proposed context-aware access management frameworks is depicted in Fig. \ref{fig.accessdiagram}. The proposed access control system considers the zero-trust context-aware system that manages and analyses the data journey from the user of medical IoT device endpoints to the cloud resources destination. 
The proposed framework is classified into three main  layers, as listed below: 
\begin{itemize}

    \item  \textbf{Cloud input sources:} Is the front-end gateway for the main input source from users, devices metadata and the context of data output either in the storage of real-time streaming.
    \item \textbf{Cloud decision engine:} Build a chain of trust for each component based on the trust scores. There are two scores: critical trust ($CT$) and bond trust ($BT$).
    Then, the engine encodes the context attributes for further analysis at a hierarchical level. In the end, it grants the final access decision, operations and constraints based on the analysis.
    
    \item \textbf{Cloud resources:}  Contains the cloud computing and storage resources that are used to process and store the metadata in the healthcare database.
    
\end{itemize}

The following subsections explain the components of the proposed framework in detail.

\subsection{Zero trust cloud input sources }

The proposed system harnesses the zero trust context-aware system to manage access from the cloud input sources. The zero trust principle is based on utilizing all available data points for access management, including user identity, location, device health, services, workload, or data classification.

\begin{figure}[H]
\begin{center} 
\includegraphics[width=10 cm]{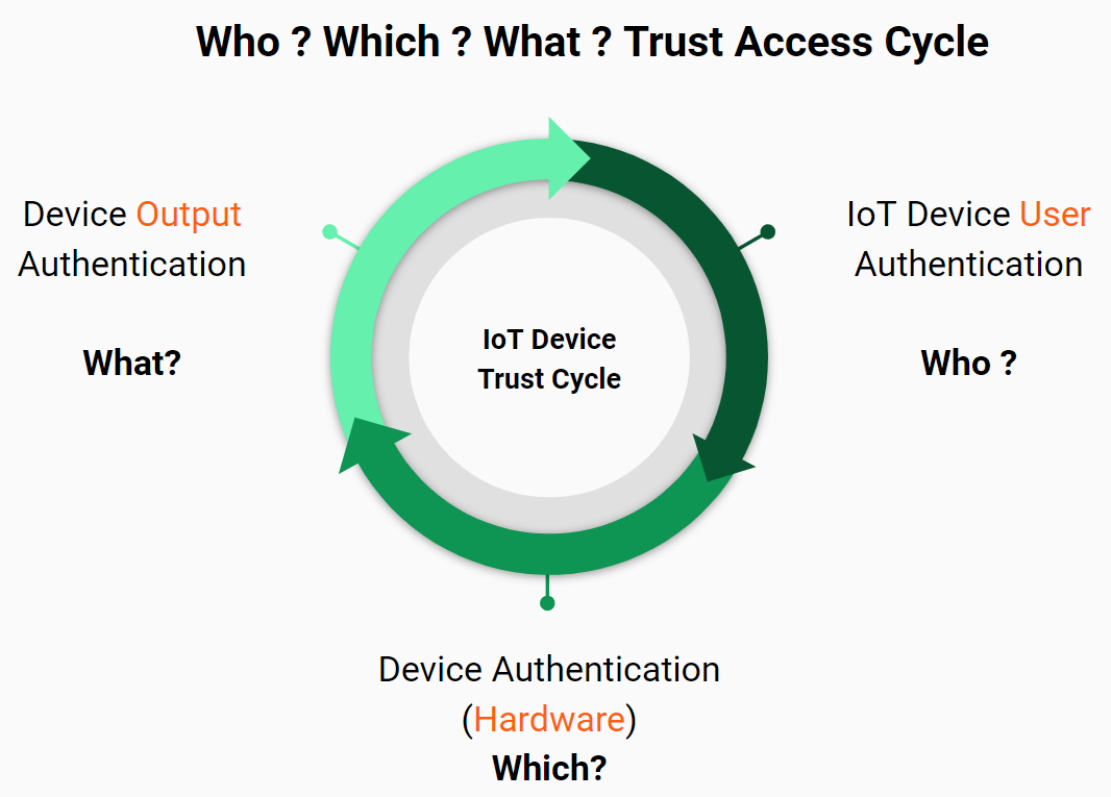}
\caption{Trust Cycle of the proposed access control framework.  } \label{fig.authenticationcycle}
\end{center}
\end{figure}

There are three main components for the context-aware cycle that consider the context of the zero-trust five elements. Fig. \ref{fig.authenticationcycle} depicts the three components of the trust cycle. Who is the user, which device is used, and what is the output? \\

The trust cycle has five elements that are pillars of the proposed zero-trust principle. \label{zt.pillars}
\begin{enumerate}
    \item User (Identity)
    \item IoT device (Hardware)
    \item Network (Device connection)
    \item Application Workload (Output patterns and scale)
    \item Data (Output transaction context)
\end{enumerate}

Where the identity relates to the user, the IoT device and the network related to the hardware component. The application workload and transaction context relate to the output component.


\subsection{ Maintain chain of trust and evaluate trust score} \label{sec.trustsore}

\begin{figure}[H]
\includegraphics[width=16 cm]{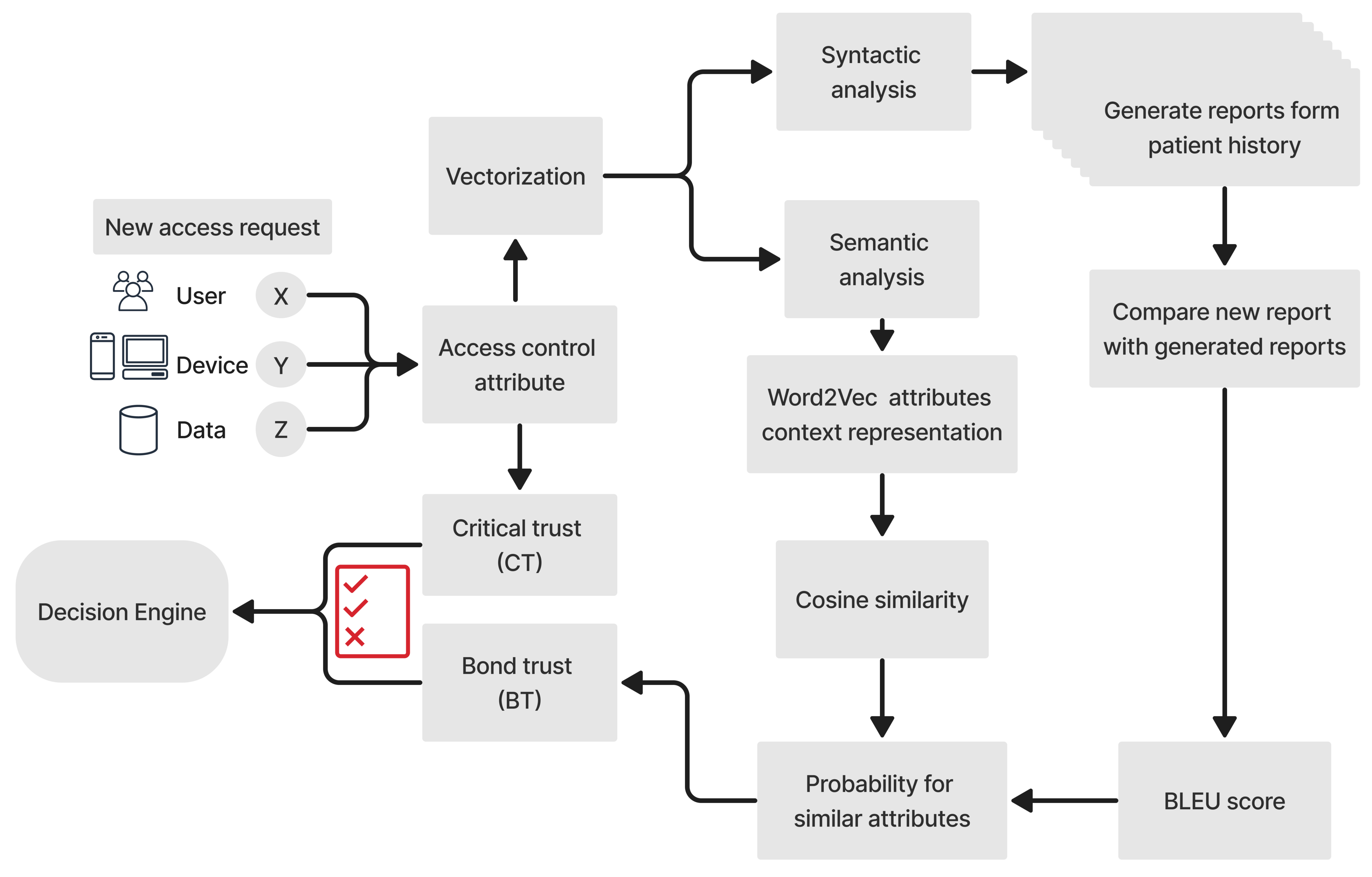}
\caption{Proposed framework of a continuous chain of trust based on the accumulated trust score of the zero trust of each access management component.   } \label{fig.chainoftrust}
\end{figure}

 While building trust to decide what type of access each component will have is essential, guaranteeing that there are no human errors or malicious attacks is challenging. To address this issue, our access control management protocol will continuously evaluate the trust score and build a chain of trust. The chain of trust is important to decide what level of access can be granted and to deny access if the connection is below the threshold of an acceptable trust score. Fig. \ref{fig.chainoftrust} illustrates the chain of trust within the cloud ecosystem.

 The proposed framework constructs two assessment scoring criteria to manage the access of distributed medical devices. First is the critical trust ($CT$), which relies on cloud-native microservices. Second is the bond trust ($BT$), which is a proposed scoring schema to manage access control, as explained below.  $BT$ uses machine learning pre-trained models to analyze the semantic and syntactic attributes from the trusted and authorized change of zero trust pillars \ref{zt.pillars}, that are related to uses, devices and data.

     \textbf{Critical Trust ($CT$):} Is the first assessment score that is used to grant initial access to the cloud ecosystem but not for direct connection to other resources. $CT$ is evaluated using cloud-based micro-services. There are four main micro-services attributes for the critical trust score. Authentication, authorization, logging and encryption are all attributes that are used to obtain a value for  \textbf{$CT$} as per Eq. (\ref{eq.CTscore}).

Each micro-service attribute is given a logical value, 1 or 0. Then, these micro-services logical values are multiplied by a scoring factor ($S_i$) based on its importance that can be set by the system admin.
   The cloud decision engine grants access status to \textbf{allow} for trusted authority, \textbf{verify} if more information is needed, and \textbf{deny} for untrusted access requests. 
   
\begin{equation}\label{eq.CTscore}
    CT=S_1\times A_1+S_2\times A_2+ S_3\times A_3+ S_4\times A_4
\end{equation}    
where $A_1$ is authentication and its scoring factor is $S_1$,  $A_2$  is the authorization and its scoring factor is $S_2$, $A_3$ is the encryption and its scoring factor is $S_3$, $A_4$ is the logging and its scoring factor is $S_4$. 
Table \ref{tabcriticaltrust} provides an example of critical trust score evaluation using different scoring factors and micro-services logical values.

\begin{table}[h]
\caption{Examples of critical trust score assessment. \label{tabcriticaltrust}}

\centering
\begin{tabular}{| p{.5cm}|p{.5cm}|p{.5cm}|p{.5cm}|p{.5cm}|p{.5cm}|p{.5cm}|p{.5cm}|p{.5cm}|p{1.5cm}|p{1.5cm}|} 
\hline
\rowcolor{lightgray}
\textbf{ID} &\textbf{$A_1$}&\textbf{$S_1$}&\textbf{$A_2$}&\textbf{$S_2$}&\textbf{$A_3$} & \textbf{$S_3$} & \textbf{$A_4$} & \textbf{$S_4$} & \textbf{Critical trust score} & \textbf{Access status}  \\
\hline

D1 & 1& 0.3 & 1 &0.4 & 1 &0.2 & 1 &0.1 & 0.9999 & Allow\\ \hline
D2 & 1 & 0.3 & 0 &0.4 & 1 &0.2 & 1&0.1 & 0.6 & Verify\\ \hline
D3 & 0 & 0.3 & 0 &0.4 & 0 &0.2 & 0&0.1 & 0& Deny\\ \hline

\hline
\end{tabular}
\end{table}

\textbf{Bond Trust ($BT$):} Once the transaction is passed the critical trust assessment, the bond trust will evaluate the relationship to other resources to build a chain of trust to the only authorized and highly trusted actors to make sure that the data or resources are accessed by designated people based on the organization policy or rules. Calculating Bond Trust is more complex and depends on different aspects.  $BT$ has two main assessment criteria. First is $BT_A$, which assesses the semantic relationship between each individual attribute stored in the health care information system. Second is $BT_B$, which assesses the syntactic relationship between the set of candidates in a generated health report.   The reason for using these two measures as the first one is that it is essential that each attribute has meaningful meaning and is related to similar attributes compared to the pre-trained ones. The second one is essential to guarantee that the attributes in the generated report are in the context of the patient's history to ensure that the report is highly likely related to the same patient and, hence, is not diagnosed falsely with the wrong case.

\begin{figure}[H]
\begin{center} 
\includegraphics[width=14 cm]{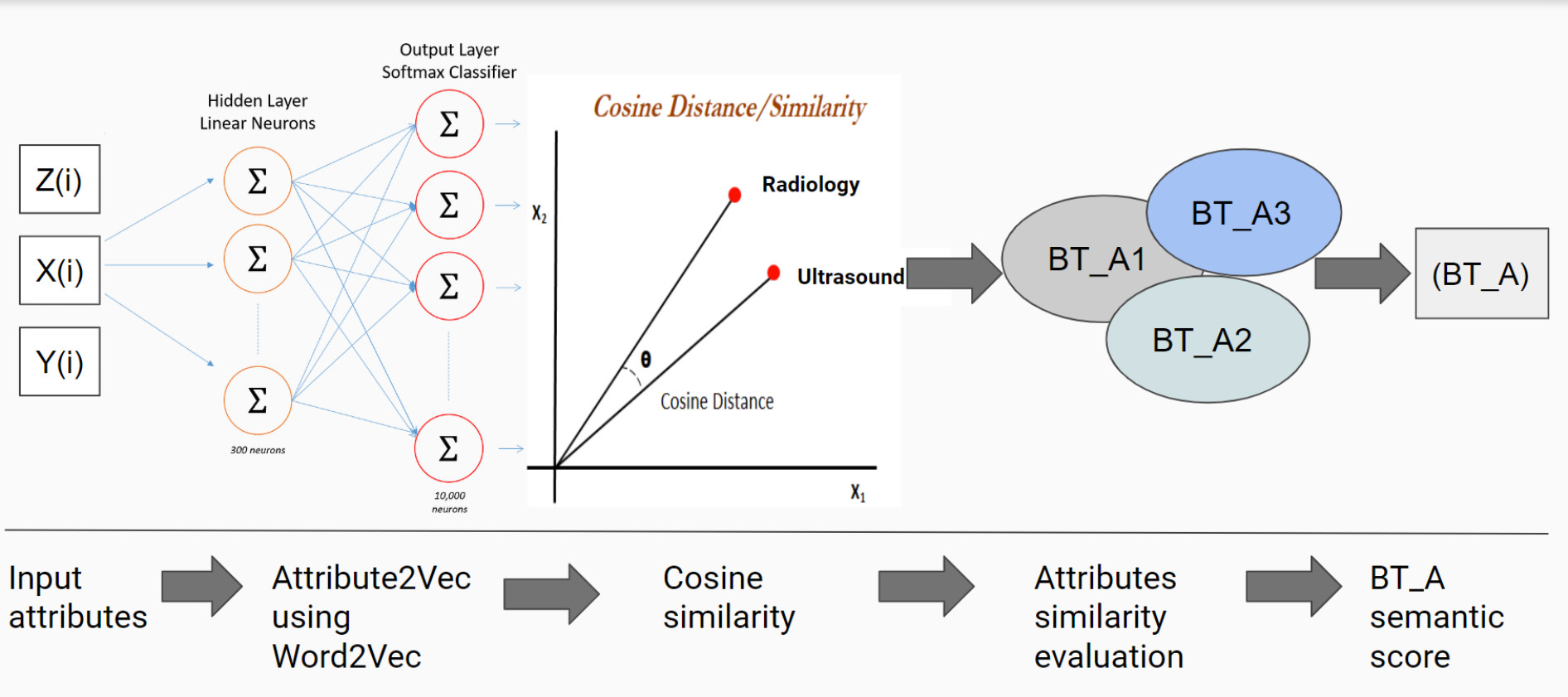}
\caption{Semantic trust assessment using attribute2vec based on word2vec model. Where $BT_{A(1)}$, $BT_{A(2)}$, and $BT_{A(3)}$ are the set of bond trusts between the three input sources $x$,$y$, and $z$. $BT$ is the final bond trust score. } \label{fig.att2vec}
\end{center}
\end{figure}

    The proposed assessment of $BT_A$ uses Attribute2Vec representation that is based on the pre-trained word2vec model \cite{mikolov2013efficient}, \cite{church2017word2vec}. The Attribute2Vec maps the attributes and their synonyms words that have the same context from the user ($x$), hardware ($y$), and output ($z$) attributes stored in their electronic health records. The skip-gram methodology \cite{hung2021word2vec} is used to derive the attributes with the same context, and we suggest in this framework using the first three words with the highest context probability. The advantage of using this assessment technique is to generalize the model by accepting a wide variety of attribute descriptions in a global context. word2vec is valid for different languages and dialects. For example, it has been used by Altibbi.com \cite{habib2021altibbivec} to train 1.5 Million medical consultation questions in the Arabic language. We recommend using a matching engine on the Vertex AI platform at Google Cloud to make sure the word embedding and vector similarity matching process is efficient and reliable. 
Fig. \ref{fig.att2vec} depicts the process of assessing bond trust.
    
    Cosine distance is used in Eq. (\ref{eq.cosineref}) to predict the similarity probability of the context of attributes of $x$, $y$, and $z$.
    
    \begin{equation} \label{eq.cosineref}
       \textit{Similarity (A,B)}=\cos (\theta ) =   \dfrac { \vec{A} \cdot \vec{B} } {\left\| \vec{A}\right\| \left\| \vec{B}\right\|} 
    \end{equation}
    where $\vec{A} \cdot \vec{B} = \sum_{i=1}^N (A_i \times B_i) $ is the dot product between two vector attributes $\vec{A}$ and $\vec{B}$. At the same time, $\left\| \vec{A}\right\| = \sqrt{\sum_{i=1}^N (A_i)^2}$, $\left\| \vec{B}\right\| = \sqrt{\sum_{i=1}^N (B_i)^2} $ are the L2-norms of the attributes $\vec{A}$, $\vec{B}$ respectively. While $\theta$ is the angle between the two vectors.

The highest probability attribute vectors between $x$, $y$, and $z$ will be used to derive the bond or semantic mutual relationship in three Bond Trust scores sets $BT_{A(1)}$, $BT_{A(2)}$ and $BT_{A(3)}$ for the relationship between $xy$, $xz$  and  $yz$. Where $BT_{A(i)}$, the bond trust sets are derived using two inputs:

\textbf{A. Cosine similarity logical evaluation:} Algorithm \ref{alg:cosine} is used to assign a logical value for the cosine similarity between two attributes. There is a given value of either one or zero based on the relationship of the attributes $x$,$y$ and $z$. The value is assigned based on the threshold of the angle $\theta$ between the two attributes. Eq. (\ref{eq.cosineref}) is used to derive $\theta$ using cosine similarity between the attribute vector product for the given ($i$) index or position for similar context attributes. The algorithm produces a set of three logical values $Sim_{A}(\vec{x}_{i}, \vec{y}_{i})$, $Sim_{B}(\vec{x}_{i}, \vec{z}_{i})$, $Sim_{C}(\vec{y}_{i}, \vec{z}_{i})$, for each given index($i$). 

\begin{algorithm}
\caption{An algorithm for proposed cosine similarity logical evaluation process}\label{alg:cosine}
\textbf{Input:} User ($x$), IoT hardware ($y$) , IoT output data ($z$), Angle Threshold ($Th_{\theta}$)\\

\begin{algorithmic}[1]
 
 \If{ ${\theta}_{xy} \geq Th_{\theta}$ }
    \State$Sim_{A}(\vec{x}_{i}, \vec{y}_{i})= 1$
    \ElsIf{${\theta}_{xy} < Th_{\theta}$}
    \State$Sim_{A}(\vec{x}_{i}, \vec{y}_{i})= 0$
    \EndIf   
 \If{ ${\theta}_{xz} \geq Th_{\theta}$ }
    \State$Sim_{B}(\vec{x}_{i}, \vec{z}_{i})= 1$
    \ElsIf{${\theta}_{xy} < Th_{\theta}$}
    \State$Sim_{B}(\vec{x}_{i}, \vec{z}_{i})= 0$
    \EndIf  
 \If{ ${\theta}_{yz} \geq Th_{\theta}$ }
    \State$Sim_{C}(\vec{y}_{i}, \vec{z}_{i})= 1$
    \ElsIf{${\theta}_{yz} < Th_{\theta}$}
    \State$Sim_{C}(\vec{y}_{i}, \vec{z}_{i})= 0$
    \EndIf\\
 \textbf{Output:} $Sim_{A}(\vec{x}_{i}, \vec{y}_{i})$,  $Sim_{B}(\vec{x}_{i}, \vec{z}_{i})$, $Sim_{C}(\vec{y}_{i}, \vec{z}_{i})$
\end{algorithmic}
\end{algorithm}

\textbf{B. Weight:} The weight is calculated using the GloVe model \cite{pennington2014glove} of word embedding to consider the co-occurrence of the attributes in a global representation context of the healthcare database. The weight is based on the conditional probability of attribute occurrence or importance as in Eq. (\ref{eq.glove}).
        
        \begin{equation} \label{eq.glove}
            w_i={\frac{P_{BA}}{P_{B}}} 
        \end{equation}
where $w_i $  is the probability of word B occurrence in the context of the word A  of a given ($i$) index of two semantic or syntactically similar attributes.


  The three scaler values of $BT_{A(1)}$, $BT_{A(2)}$ and $BT_{A(3)}$ are stored in  $BT_{A}$ as shown in Eq. (\ref{eq.BTvec}). Where $BT_{A}$ is a $1 \times 3$ vector. 
 
\begin{equation} \label{eq.BTvec}
    BT_{A}=[BT_{A(1)},BT_{A(2)},BT_{A(3)}]
\end{equation}
where $BT_1$ is the relationship score between the user ($x$) and hardware ($y$) and is derived using Eq. (\ref{eq.BT1}). $BT_2$ is the relationship score between the user ($x$) and output ($z$) and is derived using Eq. (\ref{eq.BT2}). $BT_3$ is the relationship score between the output ($z$) and hardware ($y$) and is derived using Eq. (\ref{eq.BT3}).

\begin{equation} \label{eq.BT1}
    BT_{A(1)}= \sum_{i=1}^N {(w_{i})_{xy} \cdot Sim_{A}(\vec{x}_{i}, \vec{y}_{i})}
\end{equation}

\begin{equation} \label{eq.BT2}
    BT_{A(2)}=\sum_{i=1}^N {(w_{i})_{xz} \cdot Sim_{B}(\vec{x}_{i}, \vec{z}_{i})}
\end{equation}

\begin{equation} \label{eq.BT3}
    BT_{A(3)}=\sum_{i=1}^N {(w_{i})_{yz} \cdot Sim_{C}(\vec{y}_{i}, \vec{z}_{i})}
\end{equation}
where $w_i$ is a scalar weight that is used to scale the bond score for each attribute based on the importance of the feature at given ($i$) and derived by Eq. (\ref{eq.glove}). ${N}$ is the sequence number of attributes that are numbered based on the probability of their context relationship. Only each similar class attribute of user, devices and output are multiplied by each other, and if they belong to the same category, the algorithm gives them either a 0 or 1 similarity score and then multiplies them by the scalar weight for that attribute. The step is then repeated for all attributes. The final multiplication is aggregated to have a final scalar number that resembles the combined similarity score for $BT_{A(i)}$.

$BT_{A(i)}$ vector is normalized in Eq. (\ref{eq.smax}) using the $SoftMax$ function.  The normalization process produces a new vector $BTN$ of a $1 \times 3$ dimension.  
 
\begin{equation}\label{eq.smax}
BTN_{i}=\textit{SoftMax} (BT_{A(i)}) = \frac{\exp(BT_{A(i)})}{\sum_j \exp(BT_{A(j)})}
\end{equation}
The result is stored in Eq. (\ref{eq.BTNor}) and has three scalar values that are between zero and one.
\begin{equation} \label{eq.BTNor}
    BTN_i=[BTN_1,BTN_2,BTN_3]
\end{equation}

The first part of the bond score is calculated in Eq. (\ref{eq.BTscore}) by aggregating the three normalized scores $BTN_1$, $BTN_2$ and $BTN_3$. 
\begin{equation} \label{eq.BTscore}
    BT_A=BTN_1+BTN_2+BTN_3
\end{equation} 
where $BT_A$ is between zero and one. Zero is for non-matched attributes, and one is for the highest attribute similarity match. Any number between zero and one requires an additional trust verification and reassessment.

At the same time, $BT_B$ is used to evaluate the syntactic performance of the candidate report generated from the stored data in the healthcare information system. $BT_B$ is inspired by the $BLEU$ score \cite{papineni2002bleu}, which was originally designed by IBM for machine translation scoring evaluation as shown in Eq. (\ref{eq.bleu}).

\begin{equation}
    BT_B= \min (1, exp (1- \frac{reference-length}{output-length})) (\Pi^{n}_{i=1} precision_i )^{1/n}. \label{eq.bleu}
\end{equation}
where $BT_B$ has two parts. First is the brevity penalty that compensates for the length of a short generated report. Second,  the precision for n-gram candidates. The $n$ refers to the number of candidates used to evaluate the score. $n$ typically is 4 and can be increased to include more restrictions for identifying the misuse cases. In the case of $n=4$, the $BLEU$ score needs the candidate report to match the reference template by at least four attributes.

In the case of no history, the score is zero, and it becomes more meaningful and has more weight in the case of existing patient history.  

The final bond trust normalizes the summation of $BT_A$ and $BT_B$ to keep the value between zero and 1 ad in Eq. (\ref{eq.BT}).
\begin{equation}
    BT= \frac{BT_A+BT_B}{2} \label{eq.BT}
\end{equation}

\subsection{Access control decision engine encoding and hierarchy} \label{sec.accesshierarchy}

\begin{figure}[H]
\includegraphics[width=14 cm]{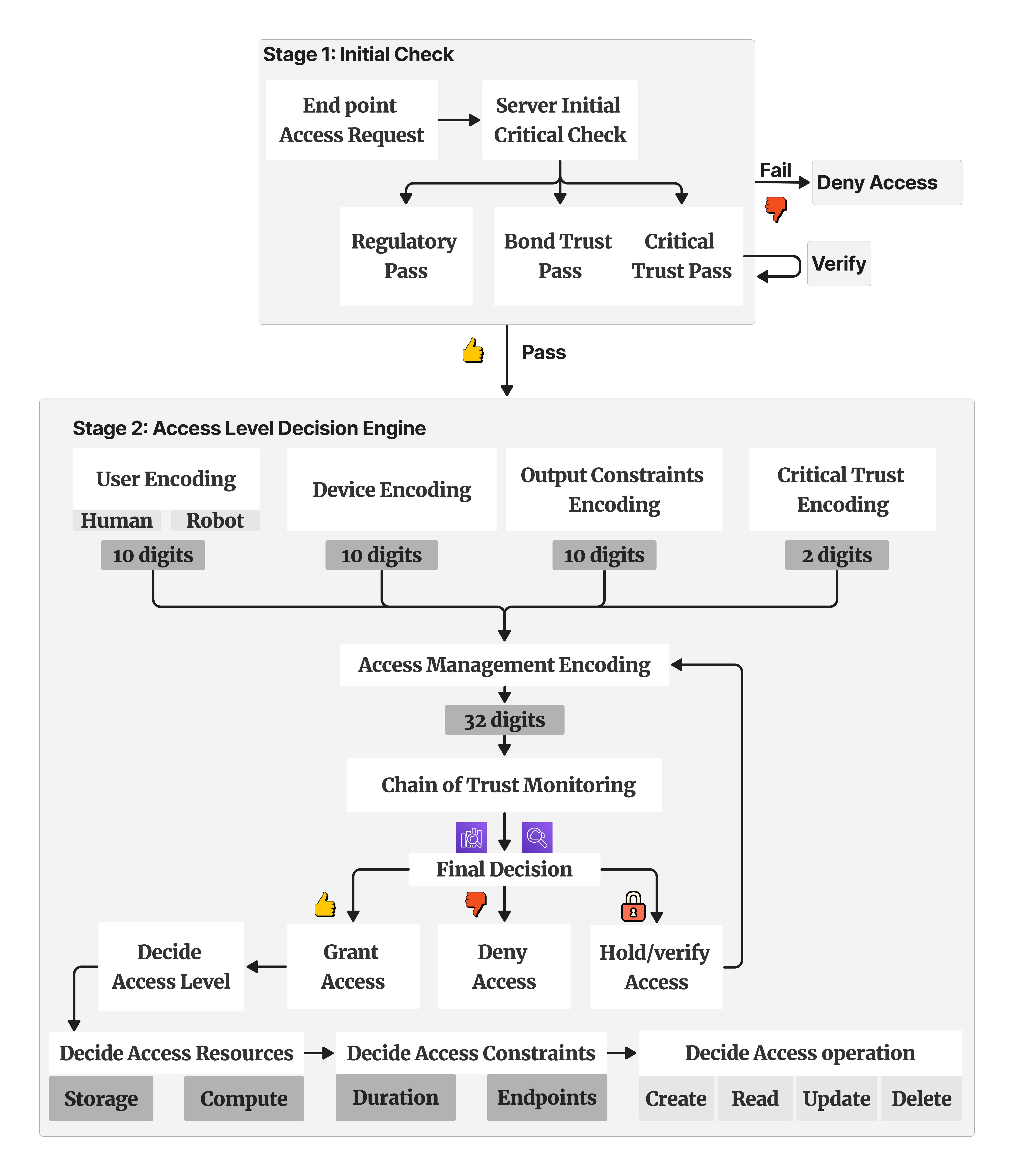}
\caption{Representative of the access control engine decision hierarchy and the encoding.  } \label{fig.accesshirerachy}
\end{figure}

There are two main stages for the decision engine encoding and hierarchy. These stages are essential to make sure that the final decision follows a chronological and logical flow of different conditions. While the encoding  Fig. \ref{fig.accesshirerachy}, depicts the two stages of encoding and hierarchy process for access control management.

\textbf{Stage one:} The decision engine performs the initial critical check for the end point device or user request access from the server. The critical check is essential to guarantee that the endpoint components have passed the regulatory compliance and critical trust score threshold, Table. \ref{tabcriticaltrust}.   An example of one of the main regulatory compliance that needs to be considered is the HIPAA, which lists 18 patient information identifiers \cite{hipaa18} that are restricted from being shared without having consent from patients and meet all security guidelines within the healthcare information system as shown in Table \ref{tabaccesshipaa}.

\textbf{Stage two:} The decision engine encodes the attributes from devices, output, users and critical trust in a 32-digit hexadecimal array. The  32-digit array is then analyzed to make the final decision. The final decision is also encoded to resemble the access level, operations, access resources and constraints, see Sec. \ref{sec.finaldeceiosn}.

\unskip

\begin{figure}[H]
\begin{center} 
\includegraphics[width=12 cm]{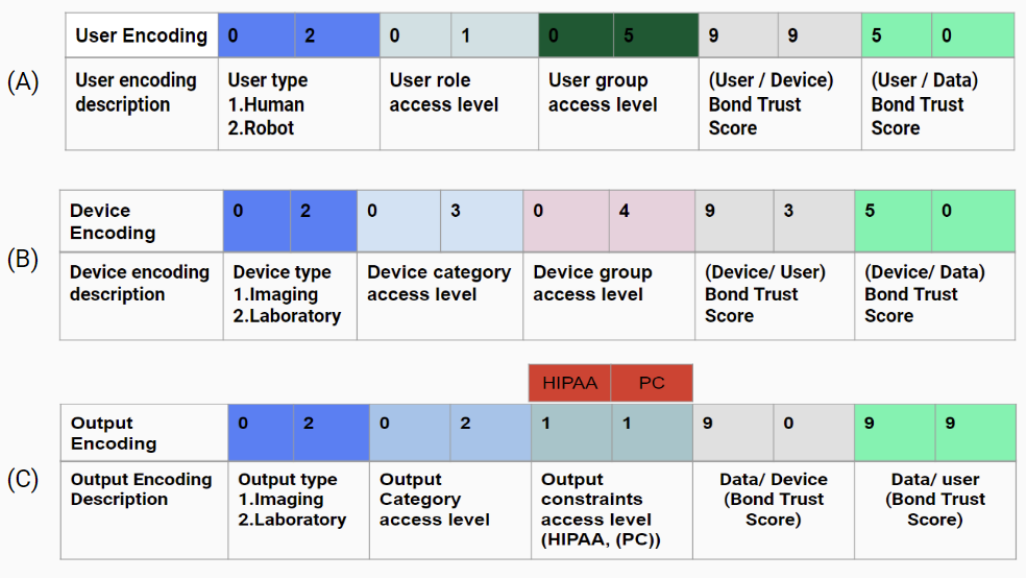}
\caption{ Proposed access control encoding. (A) User Encoding. (B) Device Encoding. (C) Output encoding. PC is the patient consent.} \label{fig.accessencode}
\end{center}
\end{figure}

Fig. \ref{fig.accessencode} illustrates an example of the encoding schema for the proposed three zero-trust components, user, device and output, based on different attributes. Each component has a 10-digit hexadecimal value and a 2-digit value for each one of the five attributes. The importance of these attributes is to ensure that the access request belongs to the designated group, has a pre-defined access level and type, and passes the bond trust.

\subsection{Final decision and access operations} \label{sec.finaldeceiosn}

The final decision has four main criteria and information, as listed below. Table. \ref{tabfinaldecision} depicts the final decision encoding information using hexadecimal digits.
\begin{itemize}
    \item \textbf{Access level:} Decide the access level for each transaction.
    \item \textbf{Access resources:} Decide which storage and computation resources will be used.
    \item \textbf{Access constraints :}  Decide what the access constraints are, such as duration, location, number of access trials, and size of data transferred. 
    \item \textbf{Access operations:} Grant access based on CRUD or HTTP method.
\end{itemize}

\begin{table}[h]
\caption{Final decision encoding. \label{tabfinaldecision}}

\centering
\begin{tabular}{| p{1cm}|p{2.2cm}|p{2.2cm}|p{7cm}|} 
\hline
\rowcolor{lightgray}
F & Decision & Hex & Encoding Description  \\
\hline

F1 & Access Level & 2 digits & To specify the five access levels. Ex. 10 for access level L0. \\ \hline
F2  & Access resources & 6 digits & The first three digits are for compute resources and the rest are for storage resources metadata.\\ \hline
F3 & Access constraints & 16 digits & 8 digits for time,  and the other eight digits for other constraints. Ex. 6421EC5F for 2023Y,03M,27D,21h,19mm,59ss. \\ \hline
F4 & Access operations& 1 digit & ex. F, in hexadecimal, for admin access of all operations\\ \hline

\hline
\end{tabular}
\end{table}

Algorithm \ref{alg:cap} shows the logical process of the proposed framework. The framework has three inputs $x$, $y$ and $z$. The initial step requires passing the threshold for $CT$ and $BT$ that is specified by the system admin. Typically $CT \geq 99.99\%$ , $BT \geq 0.7$, where each attribute in $BT_{i} \geq \theta$. If the score of $CT$ and $BT$ was zero, the access was denied, and any value between zero and threshold, the access should be verified again within a given time interval. The final access decision will be granted based on the trust scores assessment.

\begin{algorithm}
\caption{Proposed access management framework logical process}\label{alg:cap}
\textbf{Input:} User ($x$), IoT hardware ($y$) , IoT output data ($z$)\\
\textbf{Trust assessment:}  Critical trust ($CT$), Bond trust ($BT$), Trust threshold ($Th$)
\begin{algorithmic}[1]
 
 \If{$CT = 0$ , $BT = 0$}
    \State Deny  access
    \EndIf   
\While{$Th \neq 0$}
\If{$CT \geq Th$ , $BT \geq Th$}
    \State Initially accepts access
    
\ElsIf{$CT < Th$ , $BT < Th$}
    \State Verify access again
   
\EndIf\\
 \textbf{Output:} Grant final access decision
\EndWhile

\end{algorithmic}
\end{algorithm}

\section{Experiment and results} \label{sec.result}

\subsection{Dataset information} \label{dataset}
\begin{figure}[H]
\begin{center} 
\includegraphics[width=13 cm]{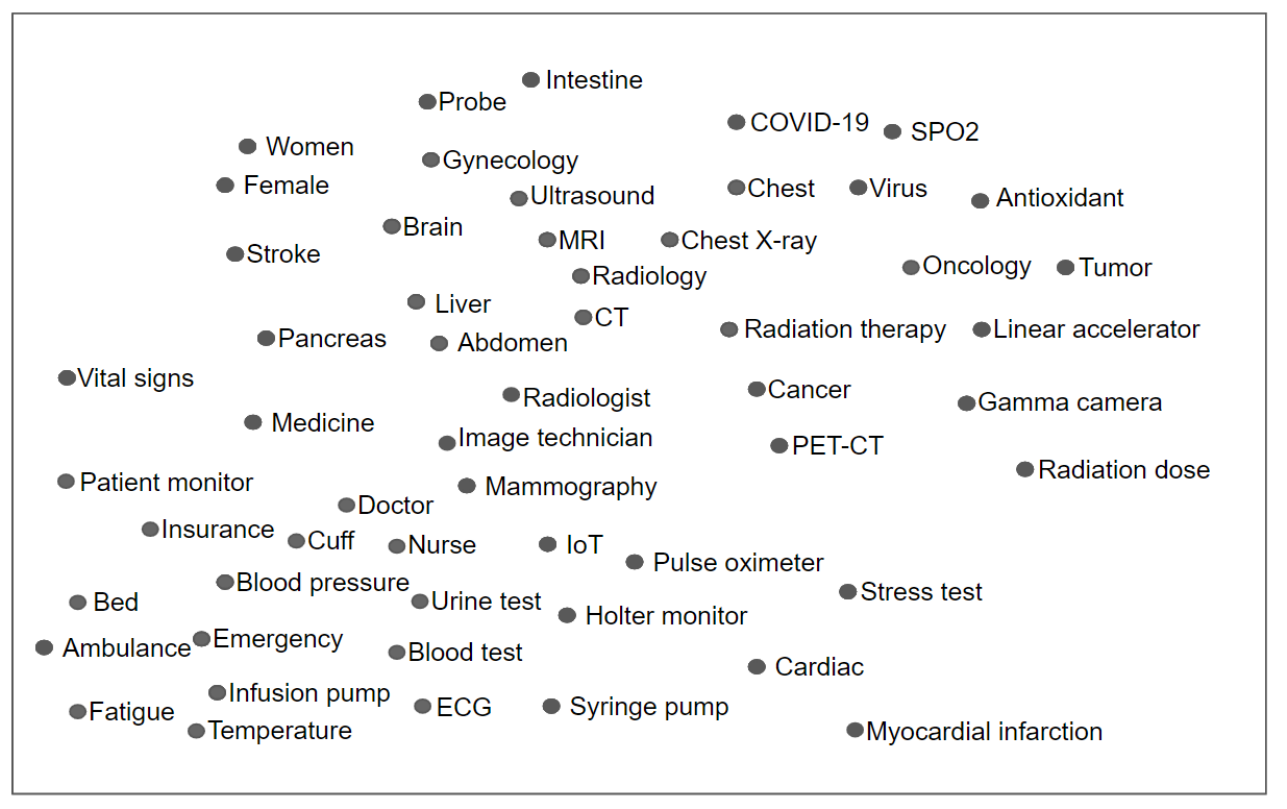}
\caption{ Example of selected attributes from the generated data using Synthea and fine-tuned the word2vec pre-trained model.} \label{fig.wr2vec}
\end{center}
\end{figure}

The dataset used in this experiment contains a synthetically generated set of attributes for users, medical IoT devices and data. The generated data is synthesized using Synthea \cite{walonoski2020synthea}. The synthetic dataset is used to fine-tune the word2vec model to enhance the result within a healthcare information system. Fig. \ref{fig.wr2vec} shows some examples of the generated dataset attributes of users, devices and data that are used to prevent misuse cases.

The synthetic dataset is not limited to only a small subset of attributes. These attributes are extracted from large language models, and they include all attributes that are related to the healthcare information system for the main three categories user, device, or data. Examples of some of these categories are listed in Table. \ref{tab:data}.

\begin{figure}[H]
\begin{center} 
\includegraphics[width=15 cm]{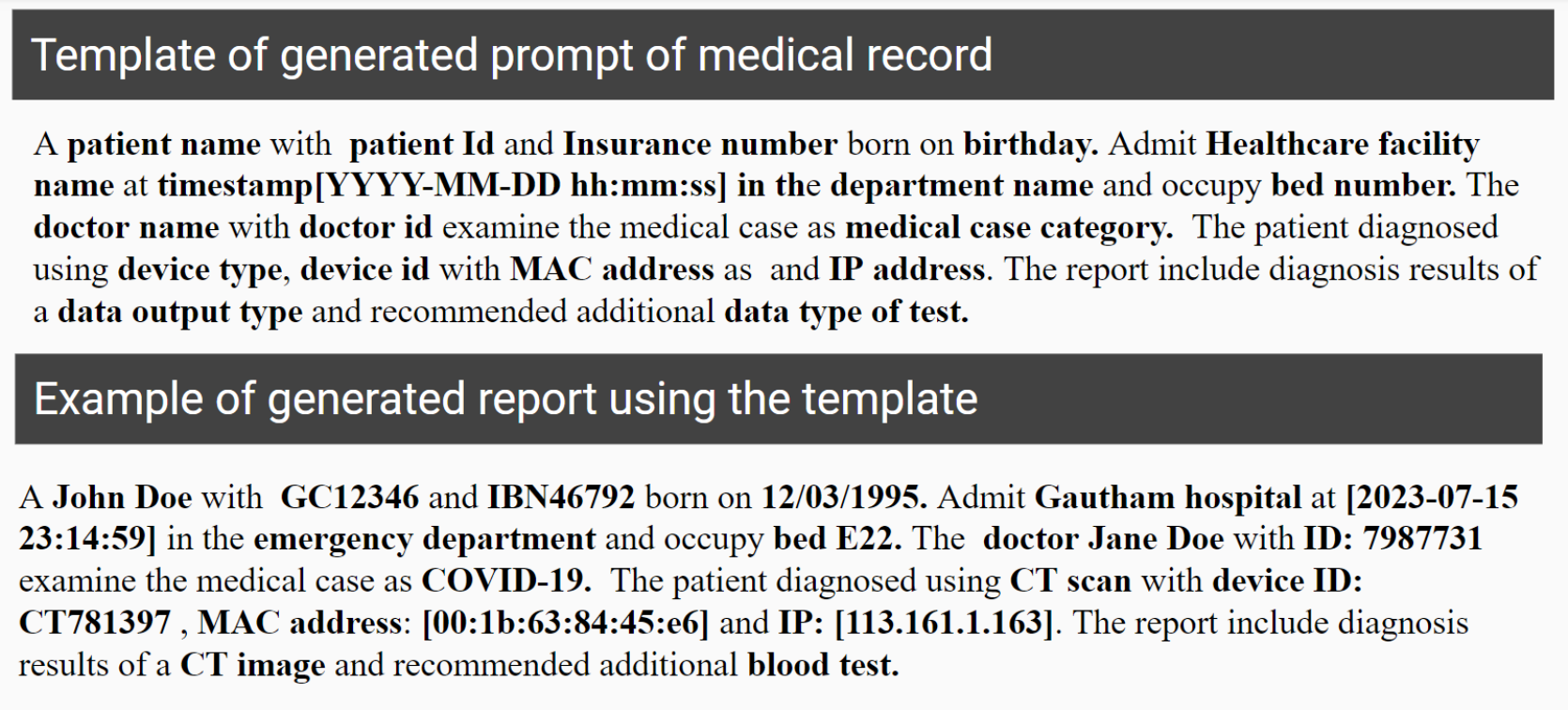}
\caption{ Arbitrary example of generated text prompt from patient history record. The mentioned names are arbitrary examples and do not refer to any true identity.} \label{fig.genAItemp}
\end{center}
\end{figure}

In order to evaluate the syntactic information, we generated a sample of different patient reports based on a predefined template from the same set of attributes mentioned before. The generated templates generate all possible syntactic and semantic similar reports that may be related to the patient based on the medical history.
Fig. \ref{fig.genAItemp} depicts the template used to generate the final report that includes information about users, devices and data. 

\subsection{Experiment results and discussion}

\begin{figure}[H]
\begin{center} 
\includegraphics[width=10 cm]{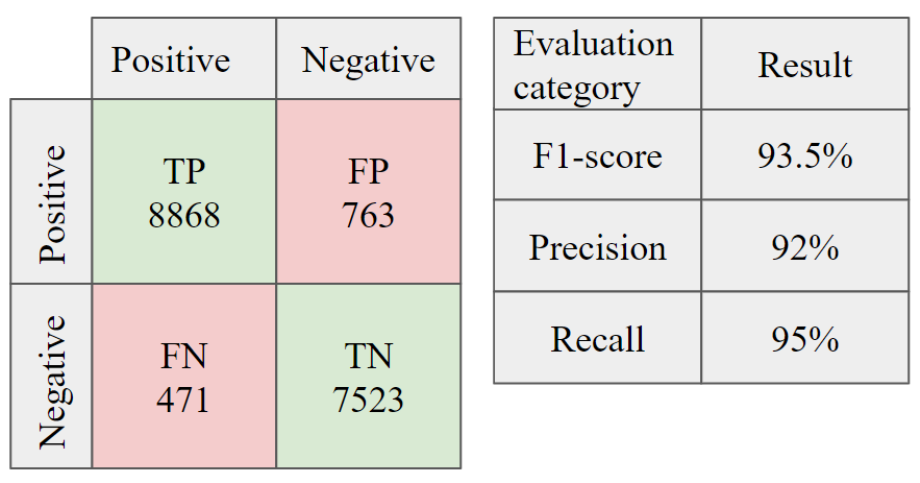}
\caption{ Confusion matrix for the ablation study on the accuracy of detecting misuse cases by identifying the relationship between selected attributes. TP is true positive; FP is false positive; FN is false negative; TN is true negative.} \label{fig.confusionmtx}
\end{center}
\end{figure}

The ablation study is examined to evaluate the accuracy of identifying the misuse cases using the proposed model by examining the relationship between different attributes based on the critical and bond trust scoring. The study was conducted using 17625 attributes for users, medical IoT devices, and data output categories. The study shows that the F1-score is $93.5\%$, which means that the proposed methodology is valid for identifying the relationship between different attributes within the healthcare information system and alleviating any misuse cases that may produce false medical reports. Fig. \ref{fig.confusionmtx} depicts the confusion matrix of the experiment result.

Evaluating the final results has different criteria for semantic and syntactic information.    Semantic information is used to evaluate the relationship of each word in the context of medical-related data. At the same time, syntactic information analysis is used to evaluate the corpus context of the newly generated report and compare it with the stored data in a healthcare information system. 
Table \ref{tab.word2vec} lists some examples of the evaluation of the $BT_{A}$ to extract semantic relationships from the healthcare information system for different medical specialty classes.


\begin{table}[h]
\caption{The precision, recall, and f1-score for evaluating $BT_{A}$ on a selected variety of speciality cases. \label{tab.word2vec}}

\centering
\begin{tabular}{| p{2.5cm}|p{1.6cm}|p{1.5cm}|p{1.5cm}|} 
\hline
\rowcolor{lightgray}
Speciality Class & Precision & Recall & F1-score   \\
\hline

Radiology & 0.880 & 0.880 & 0.880 \\ \hline
Gynecology  & 0.773 & 0.840 & 0.805 \\ \hline
Oncology & 0.793 & 0.772 & 0.782 \\ \hline
Dermatology & 0.712 & 0.740 & 0.726 \\ \hline
Cardiology & 0.833 & 0.871 & 0.851\\ \hline
Urology   & 0.765 & 0.724 & 0.744 \\ \hline
Emergency   & 0.865 & 0.834 & 0.849 \\ \hline
Dentistry    & 0.79 & 0.77 & 0.779 \\ \hline
Psychology   & 0.766 & 0.784 & 0.774 \\ \hline

\hline
\end{tabular}
\end{table}


Table. \ref{tab.assesssyntactic} shows the effect of syntactic analysis using different measures on the final decision. As shown in the table, the 1-gram measure of the accuracy of detecting one word compared to the context of reference length from the stored data in the healthcare system. While it has high precision, it is not accurate for making decisions as it does not account for the relationship with other attributes. The decision confidence increases gradually with reference to the n-gram rank as it has a more meaningful meaning.\\
The BLEU score is good for judging a corpus of attributes but performs badly on a single entry. In the case of syntactic analysis, it is more efficient to score the generated reports. However, it is not efficient for judging semantic information or detecting grammatical error sentences.\\

The proposed method got the best results as it accounts for semantic and syntactic information. The decision engine in the cloud generates different reports from the stored data that account for different synonyms, words or attributes that are related to the stored data. It also can fix any grammatical errors in the entry and suggest an attribute within the same context. This gives the method a generalized capability to assess any new report or data entry within the healthcare information system through distributed users, devices and sorted data. At the same time, the proposed method accounts for the security measure that requires authentication, authorization, encryption and logging.
Table \ref{tab.assesssyntactic} compares the confidence score of the proposed method with other scoring metrics that are used for syntactic analysis. 

\begin{table}[h]
\caption{Comparison of proposed scoring method and other metrics of syntactic analysis. \label{tab.assesssyntactic}}

\centering
\begin{tabular}{|p{2.5cm}| p{3cm}| p{2.2cm}|} 
\hline
\rowcolor{lightgray}
Metric &  Decision confidence score \\
\hline

1-gram&  26\%\\ \hline
2-gram &   33\%\\ \hline
3-gram&   47\%\\ \hline
4-gram& 66\%\\ \hline
BLEU &  71\% \\ \hline
Proposed method &  89\%\\ \hline

\hline
\end{tabular}
\end{table}


The proposed framework focuses on the cloud-AI access control system by managing the access to the cloud resources for users, devices, and data. This is done by implementing a zero-trust context-aware system that analyzes the data for each transaction and always for the uses of the users, devices, network, workload and data. The framework considers the information security model that has three pillars of data confidentiality availability. Regulatory compliance is also part of context-aware access control. HIPAA is the most important compliance that identifies protected health information data. The protected data can not be used without following a series of security and privacy protection guidelines like patient consent, disclosure agreement, de-identifications, data encryption and a well-managed access control system.\\

The framework also takes advantage of cloud-native microservices to implement critical trust assessment criteria. To build a chain of trust between different attributes for each component, the framework proposes bond trust evaluation that is inspired by the large language models.  

Table. \ref{tab.scoreass} shows a sample of results for access management decisions based on the evaluation of the critical trust ($CT$) and bond trust ($BT$) assessments.


\begin{table}[h]
\caption{Example of access management decision based on scoring evaluation. \label{tab.scoreass}}

\centering
\begin{tabular}{| p{2cm}|p{1.1cm}|p{1cm}|p{2cm}|} 
\hline
\rowcolor{lightgray}
N(Samples) & $CT$ (avg.) & $BT$ (avg.)& Decision  \\
\hline

402 & 0 & 0 & Decline \\ \hline
267 & 0.99 & 0 & Decline \\ \hline
224 & 0.99 & 0.5 & Verify \\ \hline
259& 0.99 & 0.9 & Accept \\ \hline
110 & 0.99 & 0.83 & Accept \\ \hline
479  & 0.99 & 0.79 & Accept \\ \hline

\hline
\end{tabular}
\end{table}


While the zero trust context-aware system is robust against different situations, there are different challenges and limitations to implementing it in the healthcare industry, as follows: 
\begin{itemize}
    \item \textbf{Data Privacy and Security:} ML acts as a backbone of the zero trust access control system, which requires being trained on a considerably large dataset, the size required be in millions or even billions of parameters to get efficient results. Obtaining sensitive and accurate data is challenging due to privacy concerns for health information regulatory compliance, which may limit the accuracy of the system. 
    \item  \textbf{Complexity:} The complex healthcare IT infrastructure makes it difficult to implement and manage a zero-trust context-aware access control system. These systems need to be able to integrate with existing systems and applications, and they need to be able to handle the large volume of data that is generated in healthcare settings.
    
    \item \textbf{Cost:} Implementing a zero trust access control system requires an enormous investment in backend infrastructure. The costs of implementing and maintaining these systems need to be balanced against the potential benefits, such as improved data security and reduced risk of data breaches, compared to the cost of investment.
    
    \item \textbf{Skills:} Zero trust principles rely on too many factors. These factors should be aligned with the current and most advanced technology, which requires highly skilled professionals who are always in demand due to the shortage of these skills in most employees. 
\end{itemize}

\section{Conclusions} \label{sec.conclusion}

The access control system is one of the essential components of any information system that overcomes the three main challenges of data interoperability, confidentiality, availability and integrity.
The proposed framework handled the three main challenges by proposing dynamic zero trust-based context-aware access management that analyzes the data from users, devices and output to decide the level of access, constraints, resources and operations. The results show that the combination of zero trust using cloud-based microservices and bond trust using semantic and syntactic analysis is able to detect misuse cases by a high confidence score compared to other scoring algorithms. The proposed framework also considers security and regulatory compliance in addition to semantic or syntactic information attribute search. For future work, other technical elements should be considered in the access control system design. Examples of some of these elements are quantum computing, post-quantum encryption and blockchain technology.
 \\

\newpage
\textbf{Abbreviations:}\\

\noindent 
\begin{tabular}{@{}llll@{ }}

 AI:  & Artificial intelligence \\
 AIDA:  &  Artificial intelligence and data act  \\ 
 ABC:  & Attribute-based access control \\
 BT:  & Bond trust \\
 CT:  & Critical trust \\
 CPPA:  &  Consumer privacy protection act  \\
 CML:  &  Cloud and machine learning-based access control  \\
 CRUD:  & Create, read, update, delete  \\
 DDoS:  & Distributed denial of service attacks  \\  
 GDPR:  &  General data protection regulation  \\ 
 IoT:  &  Internet of things  \\
 ML:  &  Machine learning  \\
 PIPEDA:  &  Personal information protection and electronic document act  \\   
 PIDPTA:  &  Personal information and data protection tribunal act  \\
 RBC:  & Role-based access control \\
 SQL:  & Structured query language \\

\end{tabular}

\newpage


\appendix

\section{Access control system information} \label{aceessinfo}

\subsection{ Role-based access control system important factors } 

\begin{table}[h]
\caption{Examples of Role-based access control system data sources information.\label{tabaroleinfo}}

\centering
\begin{tabular}{| p{4cm}|p{10cm}| } 
\hline
\rowcolor{lightgray}
Information source & Example of data source information factors \\
\hline

Client or operator		& User id, role, department, level of access, geographic location\\ \hline
Patient        	&   Patient ID, clinical condition, department, family doctor name, patient consent policy   \\ \hline
Resources  	&   Confidentiality, sensitivity, type of data, date ranges covered by the data, author of the data  \\ \hline
Data context  	&   System identity, transaction time, the expiration time of token data, the scope and purpose of the token, security of transaction  \\ \hline

\hline
\end{tabular}
\end{table}







\subsection{ HIPPA 18 Restricted Patient Identifiers} 
\begin{table}[h]
\caption{HIPAA 18 patient protected identifiers \cite{hipaa18}. \label{tabaccesshipaa}}

\centering
\begin{tabular}{| p{1cm}|p{6cm}|p{1cm}|p{6cm}|} 
\hline
\rowcolor{lightgray}
Iden. & Description & Iden. & Description  \\
\hline

1 & Names & 10 & IP address \\ \hline
2 & SN (Social security number) & 11 & Medical records number\\ \hline
3 & Geographic locations smaller than states & 12 & Biometrics identifiers \\ \hline
4 & Telephone numbers & 13 & Health plan beneficiary numbers \\ \hline
5 & Fax numbers & 14 & Full face photographs \\ \hline
6 & Devices IDs and serial numbers & 15 & Account numbers \\ \hline
7 & Email address & 16 & Any other unique identifying numbers \\ \hline
8 & Web URLs & 17 & Certificate; license numbers \\ \hline
9 & Vehicle identifiers (eg. license plate) & 18 & All element of dates (Except years) \\ \hline

\hline
\end{tabular}
\end{table}

\subsection{Data category example}

\begin{table}[h]
\caption{Example of some attributes from the generated dataset categories. \label{tab:data}}

\centering
\begin{tabular}{| p{8cm}|p{2cm}|} 
\hline
\rowcolor{lightgray}
Attribute example & Category  \\
\hline

Position & User \\ \hline
Department & User  \\ \hline

Speciality & User  \\ \hline
ID & User \\ \hline

Insurance number & User \\ \hline

User access level & User \\ \hline
User consent & User \\ \hline

Password & User  \\ \hline
Manager ID & User  \\ \hline

MAC address & Device  \\ \hline
IP address & Device  \\ \hline

Device model & Device  \\ \hline
Device type & Device  \\ \hline

Device location & Device  \\ \hline
Device manufacture & Device  \\ \hline

Data category & Data  \\ \hline
Data encryption & Data \\ \hline

Data storage location & Data  \\ \hline
Storage type (Ex. Container, SSD, VM...) & Data \\ \hline

Data sensitivity level & Data  \\ \hline
Data compliance & Data \\ \hline

\hline
\end{tabular}
\end{table}






\newpage

 \bibliographystyle{elsarticle-num} 
 \bibliography{cas-refs}





\end{document}